# MonTrees: Automated Detection and Classification of Networking Anomalies in Cellular Networks

Mohamed Moulay, Rafael Garcia Leiva, Pablo J. Rojo Maroni, Vincenzo Mancuso, Antonio Fernandez Anta, Ali Safari Khatouni

*Abstract*— **The active growth and dynamic nature of cellular networks makes network troubleshooting challenging. Identification of network problems leveraging on machine learning has gained a lot of visibility in the past few years, resulting in dramatically improved cellular network services. In this paper, we present a novel methodology to automate the fault identification process in a cellular network and to classify network anomalies, which combines supervised and unsupervised machine learning algorithms. Our experiments using real data from operational commercial mobile networks obtained through drive-test measurements as well as via the MONROE platform show that our method can automatically identify and classify networking anomalies, thus enabling timely and precise troubleshooting actions.**

*Index Terms*—**Network anomalies, mobile measurements, feature selection, clustering, decision trees, machine learning.**

## I. INTRODUCTION

THERE has been a remarkable evolution in cellular networks during the recent years. With 4G networks, and even more with the recently roll out of 5G, network services have gained a large degree of intelligence, and involve intensive access to both data communication and computing resources. With the evolution of cellular networks, it has also come an increase in structural complexity and heterogeneity of services, which requires the constant monitoring of the communication system. Indeed, the early detection and correction of operational issues and malfunctioning components in the network is needed to provide network customers with flawless quality of service (QoS) [1]. However, the development and deployment of monitoring subsystems have to face the fast increase in technical complexity of networks [2], and a steady increase in the number and capabilities of mobile devices, hence in the number and complexity of service instances requested to the network [3]. To deal with these phenomena, operators are investing resources into the automation of the maintenance and troubleshooting tasks through self-healing functionalities within the scope of intelligent self-organizing network operation tools. Self-healing network mechanisms are accountable for detecting, identifying, and making decisions on recovery actions [1].

There exist various proposals for making fault detection and self-healing systems effective in mobile networks [4]. However, while traditional approaches lack flexibility and do not scale, newly-defined approaches based on machine learning (ML) lack *interpretability* of results, which hinders the triggering of proper and effective troubleshooting actions when a system fault is detected.

In this work, we join the ML research stream while focusing on the automated detection and classification of possible network performance anomalies. For training and model evaluation, we use real operational network data collected for cellular service auditing purposes by Nokia in various European countries, and we complement them with real operational data gathered by means of the MONROE platform, operating in several European countries as well [5]. Differently from existing proposals, we develop a methodology around an interpretable, cost efficient, scalable, and accessible combination of supervised and unsupervised ML algorithms.

### A. Related Work

Early works on fault detection suggested the use of time series *regression* methods and *Bayesian networks*. For instance, Khanafer *et al.* [6] proposed a method based on Bayesian networks to detect faults in cellular systems, in which they apply different algorithms to discrete performance metrics. Other works, such as [7], rely on a scoring-based system, in which the authors build the fault detection subsystem around labeled fault cases. These cases were previously identified by *experts*, using a scoring system to determine how well a specific case matches each diagnosis target. The work presented in [8] is based on a supervised genetic fuzzy algorithm that learns a fuzzy rule base and, as such, relies on the existence of labeled training sets. Indeed, most of the techniques proposed in the literature focus on using supervised machine learning algorithms [6]–[8]. In this paper we show that it may be convenient to combine different supervised and unsupervised techniques, to achieve interpretability of results, among other features.

Other works make use of advanced mathematical and statistical tools. For instance, Ciocarlie *et al.* [9] address the problem of checking the effect of network changes via monitoring the state of the network, and determining if the changes resulted in degradation. Their fault detection mechanism uses *Markov logic networks* to create probabilistic rules that distinguish between different causes of problems. A framework for network monitoring and fault detection is introduced in [10], using *principal component analysis* for dimension reduction, and kernel-based semi-supervised fuzzy clustering with an adaptive kernel parameter. To evaluate the algorithms, they use data generated by means of an LTE system-level simulator. The authors claim that this framework proactively detects network anomalies associated with various fault classes. These methods lack the flexibility of ML-based ones and, differently from our proposal, cannot be fully automated for a generic network context.

Notably, observing the literature, it is worth to mention that most of the proposals are evaluated only through network simulators, and that existing proposals help to detect network issues but they do not help to interpret the network behavior. In contrast, in our work we use data collected in real operational networks, and propose a fully automated ML-based methodology that leads to an easy interpretation of network behaviors. This includes identifying not only the occurrence of problems, but also their root cause.

### B. Original Contribution of the Work

By means of studying the behavior of real networks with respect to TCP performance, the main contribution of this article is a comprehensive ML-based complex methodology that ($i$) identifies if a network is behaving as expected or is under-performing, and ($ii$) automatically determines with high accuracy the root causes that lead to performance issues. In addition, we provide an open source implementation of our methodology, which is based on the use of the Scikit-learn library for Python [11]. Besides, we use valuable real data from commercial networks to test our proposal. This work significantly extends our workshop publication [12], in which we presented a preliminary version of our methodology. In there, we only focused on TCP throughput anomalies and rely on a subset of the Nokia dataset used in this manuscript.

### C. Organization of the Paper

The structure of the rest of the paper is as follows. Section II presents background on the ML algorithms used to develop the methodology. Section III describes out measurement platform and the datasets we use. Section IV takes a detailed look into the methodology we propose. Section V illustrates the implementation of our methodology and analyzes the results it achieves when using real data. We summarize and conclude the paper in Section VI.

## II. BACKGROUND ON MACHINE LEARNING

Recent advances in ML are showing its potential use in many different application areas [13]. Networking is one of these areas that could greatly benefit from ML, for instance for the detection and classification of anomalies, or the prediction of its future performance. In this paper we use two well-known and solid ML techniques. The first technique is decision trees, which is used for supervised classification (i.e., it requires a properly labelled dataset from which it learns). The second technique is $k$-means, which is used for unsupervised clustering. Instead of using these techniques to blindly classify and cluster data, we use them in tandem, to make it possible to understand the features of the available dataset (containing time series of network and TCP attributes), and so that the models and results they obtain are interpretable by domain experts. We avoid using Deep Neural Networks (DNN) since they require lots of data, while we target the design of a methodology that incurs limited costs for running network measurements (e.g., rely on a limited number of drive tests).

### A. Decision Trees

A decision tree is a classification approach that assigns a class to a data item by comparing its attributes with certain splitting values for these attributes [14]. This is done in the form of a tree, so each internal node of the tree has an attribute and a splitting value, which are used to send a data item to one of the two children of the node. Each leaf of the tree is associated to a class, so a data item is assigned the class of the leaf it reaches after traversing the tree starting at the root. A nice property of decision trees is that they are interpretable: it is clear from the path followed why a given data item has been assigned a given class.

A decision tree is built from a pre-classified training dataset, which is used to choose the attributes and splitting values of internal nodes, and the classes assigned to the leaves. There are several ways of choosing attributes and splits at each point of the tree construction. In this paper we use the extensively adopted Classification And Regression Tree (CART) algorithm [15] for this, which uses the *Gini impurity* metric [13] to choose the best attribute and split at each point. The Gini impurity value, denoted as $\gamma$, for a set of items belonging to $k$ classes is computed as

$$\gamma = \sum_{i=1}^{k} p_i(1 - p_i),$$

where $p_i$ is the fraction of items in the set that belong to class $i$. The Gini impurity ranges between 0 and 1, where 0 indicates that all items belong to a single class.

### B. $k$-means

$k$-means is an unsupervised method for finding compact clusters (and their corresponding cluster centroids) in a set of unclassified data items [13]. The parameter $k$ of the method is the number of clusters to be found. The $k$-means algorithm partitions the data into $k$ subsets so that the distances between the data items and the centroid of their subset is minimized.

Mathematically speaking, this algorithm aims at finding the partition $\{S_1, \ldots, S_k\}$ of the dataset that minimizes the following sum-squared error function:

$$C = \sum_{i=1}^{k} \sum_{x \in S_i} \|x - c_i\|^2,$$

where $\|x - c_i\|$ is the Eucledian distance between item $x$ and $c_i$, and $c_i$ is the centroid of all the items in $S_i$.

## III. MEASUREMENT PLATFORM AND DATASETS

In this section, we present the measurement platforms and the passive probe used to extract flow level statistics from TCP flows.

### A. MONROE measurements and Tstat

MONROE [16] is an open-access platform for multihomed experimentation in heterogeneous mobile broadband (MBB) networks. MONROE consists of a large set of nodes, both mobile (e.g., aboard of public transport vehicles) and stationary (e.g., university campus), all multihomed with three operators using commercial mobile carriers. MONROE enables controlled and repeatable end-to-end measurements that are crucial for evaluating MBB networks performance. Other experimental platforms exist, such as CAIDA Ark [17], PlanetLab [18], or RIPE Atlas [19]. However, those platforms are limited to wired broadband networks and are not multihomed. MONROE enables users to run custom experiments by means of Docker containers[1] and to schedule experimental campaigns to collect data from operational MBB and WiFi networks, together with physical layer information (and metadata, in general).

MONROE is complementary to crowdsourcing approaches and the control over the measurement environment tackles the shortcomings of crowd data (i.e., repeatability and privacy issues), though at the cost of a smaller geographical footprint. In a nutshell, MONROE has five important propertied which makes it unique for measuring MBB networks. These properties are large scale and diversity, mobility, fully programmable nodes, multihoming support, rich context information, and easy to use platform [20]. MONROE has 150 nodes in 4 countries in Europe (Italy, Norway, Spain, and Sweden). The node has a small programmable computer and supports 4 interfaces: three cellular modems and one WiFi modem. There are 95 mobile nodes out of 150 total nodes to operate on public transport vehicles (streetcars, buses, and trains) and on delivery trucks. Each node has a Linux Debian "stretch" distribution[2].

MONROE nodes are instrumented with a passive probe, Tstat[3], to sniff raw packets while they are passing through all network interfaces. Tstat captures traffic at flow level; it processes packets as they are captured and extracts information about layer-3 and layer-4 flows [21]. Tstat also brings traffic classification capabilities through behavioral classifiers [22], high-level visibility on encrypted traffic through the analysis of Domain Name System (DNS) queries [23], and a thorough characterization of activities in the monitored network.

Tstat provides a valuable set of statistics, over which are common to all flows, e.g., source and destination IP addresses, timestamp of the first and last packet seen, number of bytes and packets exchanged, and connection duration. Other statistics instead depend on the layer-4 protocol. While for UDP only the source and destination port numbers are reported, TCP statistics have more than 100 different metrics such as counters for TCP flags, i.e., SYN, ACK, FIN, RST, timestamps for first and last packet with payload, number of retransmitted bytes and packets, etc.

### B. Nokia drive-test measurements

To benchmark quality of service in mobile networks, continuous drive tests with end-to-end test scenarios are performed every day internally by the network operators (Quality Teams), and externally by third-parties or government regulators. Each of these test campaigns can have up to tens of thousands of individual test cases, from which specific metrics are calculated. These drive tests are normally executed with off-the-shelf testing equipment (NEMO, TEMS, Swissqual, etc.), capable of running predefined sequences of tests and collecting relevant low level radio and traffic information, as well as application performance statistics. The dataset used for this experimental validation is a real Test Record (TR) used by Nokia for the assessment of various mobile networks in year 2019. The drive-test datasets used in this study have been generated by processing all the information provided by the testing equipment and aggregating it at test level (count, sum, min, max, average, percentiles, etc.). The resulting datasets have a single row per test and hundreds of columns summarizing all the dimensions (date, time, location, network element information, etc.) and features (radio, TCP/IP, application, etc.) related to that particular test. The data transmitted in the drive tests is synthetic and does not include any customer's sensitive information, hence respecting the European Union General Data Protection Regulation (GDPR) and similar regulations (i.e., the data has been generated by a testing device and not by real users). Finally, potentially commercial sensitive data, such as the identity of the network operator, has been anonymized.

### C. Datasets Used in the Study

Here, we present a quantitative overview of the datasets we use in this paper.

Table I reports a breakdown of the data collected with MONROE for the services used for this study for all operators[4] available in their respective countries. We divide experiments into mobile vs wired connections. Overall, We have three mobile network operators and a wired connection for each node, although we only use one operator at a time to avoid interference and overloading the node's board. For the experiments analyzed in this paper, we consider the historical data collected by several MONROE nodes in four European countries over several months. Each node performed different experiments, which caused them to contact Facebook, Twitter, YouTube, and Google. The data collection duration varies from 19 months to 23 months, depending on the country and service, and the node type (static or mobile) used. The database with the services used in this extension for all countries and operators totals 29.2 GB in CSV format, with 144 attributes comprising millions of TCP flows.

Regarding the Nokia's drive-test dataset, we have 2.2 GB in CSV format, containing 358 514 rows of data and 1164 attributes obtained from direct measurements and post processing of pcap files, which produced several conditional statistics on the selected performance indicators. (The pcap themselves are not included in the database). For instance, the drive-test dataset reports separate statistics for the throughput and RTT observed over the entire download of a file or over the initial $n$ seconds, for multiple values of $n$. This dataset is also rich in terms of metadata, which allows us to filter experiments by test type, infrastructure, operator, vendor, device type, and communication technology, among other parameters.

---

[1] https://www.docker.com
[2] https://wiki.debian.org/DebianStretch
[3] http://tstat.polito.it/

[4] We do not report the name of the operator for the sake of privacy.

TABLE I: An overview of the dataset collected in Norway, Sweden, Italy, Spain.

| # Services | Countries | # Nodes | # Months | Size (GB) |
|---|---|---|---|---|
| Facebook | all | 239 | 24 | 11.0 |
| Google | all | 240 | 24 | 10.0 |
| Youtube | all | 239 | 19 | 5.1 |
| Twitter | all | 231 | 18 | 3.1 |

## IV. METHODOLOGY

In this section, we present the methodology that we propose to detect and classify networking anomalies at radio access and transport layer. Our methodology exploits the Throughput Data Rate (TDR), i.e., the TCP goodput, as a well-defined key performance indicator (KPI) to characterize anomalies in a dataset reporting TDR and several radio and transport-level observations ("data attributes"). We also leverage the fact that the TDR is strongly correlated to the Round Trip Time (RTT) [24] to build a prediction model.

Fig. 1 shows the four steps of the proposed methodology, commencing with i) processing the target KPI to classify available data points according to KPI ranges (this is our "ground truth"). Then we ii) build a classifier for the data points solely based on RTT data attributes to predict TDR values (this is therefore a model), so as to be able to compare the output of the classifier with the ground truth, which leads to detecting anomalies. Thereafter we iii) cluster detected anomalies based on selected attributes (related to either radio or transport layers), which allows to finally iv) classify anomalies based on their cause(s).

### A. Target Variable Characterization

The first step of the methodology (see Fig. 1) starts by characterizing the target variable, the TDR, using a statistical percentiles approach which represents our ground truth on the quality of the TDR. In practice, the dataset is split into three groups: data points with TDR values above the 90-th percentile (Good TDR samples), below the 10-th percentile (Bad TDR samples), and everything else (OK samples). Regulators and self quality assessment teams usually adopt this approach in the analysis of a complex system; in our case, the Nokia team involved in the measurements recommended this approach. Of course, other possible percentile thresholds are possible; for instance, the 20-th and 80-th percentiles could be used. This choice does not affect the proposed methodology.

### B. Detecting Anomalies

After dividing our target variable KPI (TDR), we turn the attention to an interpretable classifier that can classify the TDR into three classes (i.e., Bad, OK, or Good). There exists a wide range of classifiers fit for this job, but in this particular case, we are looking at an interpretable classifier; for instance, decision trees are humanly interpretable (white box) [12]. Thus, we use all the input data to build a decision tree with CART, which classifies the data items into these classes (Bad, OK, and Good TDR), using only the RTT data attributes (e.g., the average/maximum/minimum/standard-deviation RTT observed). We do so since our datasets contain data about webpage visits, for which it is known that the TDR almost directly originates from the observed RTT [24]. The depth of the tree is limited to $\lfloor (\log_2 n)/2 \rfloor$, where $n$ is the number of samples. Thus rule allows to easily and automatically accommodate different dataset sizes into our methodology.

As a result of the above process, we have a tree that classifies correctly a large portion of the data items. Intuitively, these are experiments in which the values of TDR and RTT are consistent. However, some items are not correctly classified by the tree. We consider these as *anomalies*, since data attribute (RTT) does not explain the target (TDR). For instance, in the next section, we will observe file download and web page experiments in which TDR and RTT values are inconsistent. We want to explore these data items further, since they may be symptoms of an underlying network problem.

### C. Clustering Anomalies

In the third step of our methodology, we restrict our attention to the anomalous data items misclassified by the initial decision tree. The general idea is to apply a generic clustering technique (i.e., $k$-means) leveraging Radio data attributes (e.g., Received Signal Strength Indicator (RSSI), Reference Signal Received Power (RSRP), Reference Signal Received Quality (RSRQ)) and TCP data attributes (e.g., congestion window size, receive window size, packet lost values, TCP idle-time) separately, and observe if these attribute combinations yield any groups. An expert handpicked these attributes since the idea is that what can not be explained within the relation between the TDR and RTT is an anomaly that we should investigate through other data attributes.

In general, we have to identify a number $c$ of potential causes for the misclassification, and for each cause, a set of data attributes that can be used to characterize the problem. As mentioned we use $c = 2$ here and, in the next section, we identified potential anomalies caused by *TCP problems* and *radio problems*. Hence, using only the data attributes that correspond to each potential cause, clustering is used to divide the anomalous data items into two clusters per cause.

By applying this process for each of the $c$ potential causes of problems, we classify anomalous data items into $2^c$ classes.

### D. Classifying Anomalies

The final step involves building a second decision tree with the full collection of data attributes (e.g., Radio and TCP) identified and trained using the items clustered with $k$-means in the previous step. The class into which this second tree classifies a data item reflects what makes it to be anomalous, and it is one of the possible $2^c$ classes identified. We use Gini as the metric to determine the best Radio and TCP data attributes to split the new classes (Radio OK/TCP OK, Radio OK/TCP Problem, Radio Problem/TCP OK, Radio Problem/TCP Problem) at each point. For instance, in the next section, the tree obtained in this final step determines if a given TCP flow is anomalous because of TCP problems or radio problems (or both, or none, see Fig. 2). Observe that the outcome may show that there are several causes for the same data item. It is also possible that no cause is assigned to a data item, maybe because it is a false positive or because the actual cause is not among the set of $c$ considered causes (TCP or Radio, in our case).

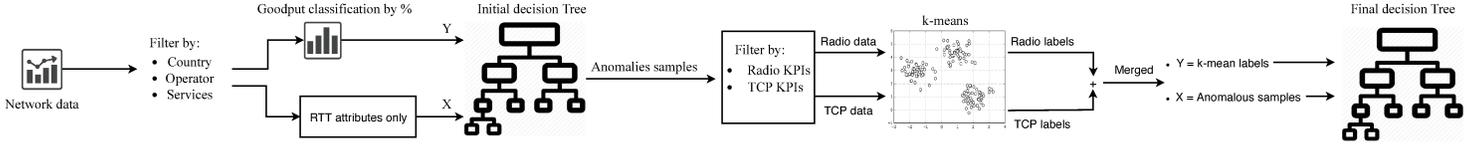

Fig. 1: Summary of the methodology starting with the characterization of the target variable percentile wise, initial classification using the RTT as the data attribute of choice, grouping anomalies with k-means (with Radio and TCP related data attributes), and final classification of anomalies with a decision tree.

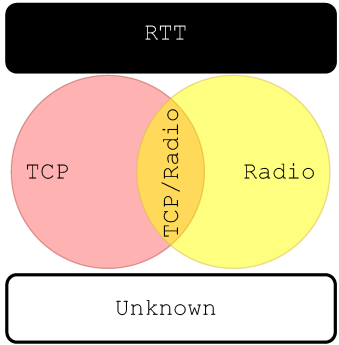

Fig. 2: Possible causes of a performance anomaly, starting with the RTT as the first choice and moving downwards to identify other factors that cause anomalies in the TDR observed (when using TCP for downloading files). The conclusion may be that the cause of the observed anomalies is not identifiable given the current model.

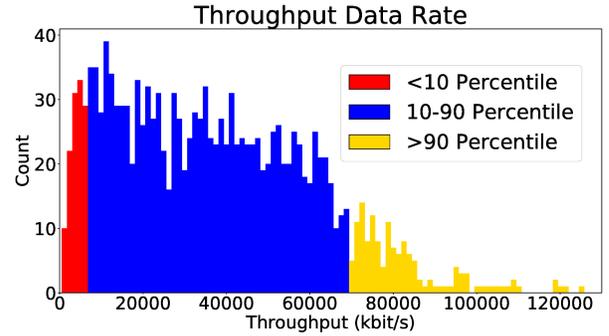

Fig. 3: Distribution of Throughput Data Rate.

TABLE II: Summary of the decision tree rules of Fig. 4 producing the most significant RTT attributes against the TDR percentile class split shown in Fig. 3

| Rules | Prob. | Class |
|---|---|---|
| If Abs_RTT_VolStep_630KB <= 225.468 | 0.84 | TDR OK |
| If Abs_RTT_VolStep_630KB <= 225.468 and If Abs_RTT_max <= 130.5 | 0.58 | TDR OK |
| If Abs_RTT_VolStep_630KB <= 225.468 and If Abs_RTT_max <= 130.5 and if Abs_RTT_VolStep_240KB <= 37.104 | 0.54 | TDR Good |
| If Abs_RTT_VolStep_630KB <= 225.468 and If Abs_RTT_max <= 130.5 and if Abs_RTT_VolStep_240KB >37.104 | 0.70 | TDR OK |
| If Abs_RTT_VolStep_630KB <= 225.468 and If Abs_RTT_max >130.5 | 0.92 | TDR OK |
| If Abs_RTT_VolStep_630KB <= 225.468 and If Abs_RTT_max >130.5 and If Abs_RTT_max <= 1130.0 | 0.94 | TDR OK |
| If Abs_RTT_VolStep_630KB <= 225.468 and If Abs_RTT_max >130.5 and If Abs_RTT_max >1130.0 | 0.56 | TDR Bad |
| If Abs_RTT_VolStep_630KB >225.468 | 0.82 | TDR Bad |
| If Abs_RTT_VolStep_630KB >225.468 and If Abs_RTT_VolStep_240KB <= 93.478 | 0.60 | TDR OK |
| If Abs_RTT_VolStep_630KB >225.468 and If Abs_RTT_VolStep_240KB <= 93.478 and If Abs_RTT_max <= 1467.5 | 0.9 | TDR OK |
| If Abs_RTT_VolStep_630KB >225.468 and If Abs_RTT_VolStep_240KB <= 93.478 and If Abs_RTT_max >1467.5 | 1.00 | TDR Bad |
| If Abs_RTT_VolStep_630KB >225.468 and If Abs_RTT_VolStep_240KB >93.478 | 0.89 | TDR Bad |
| If Abs_RTT_VolStep_630KB >225.468 and If Abs_RTT_VolStep_240KB >93.478 and If Abs_RTT_max <= 1113.5 | 0.74 | TDR Bad |
| If Abs_RTT_VolStep_630KB >225.468 and If Abs_RTT_VolStep_240KB >93.478 and If Abs_RTT_max >1113.5 | 0.96 | TDR Bad |

In our methodology, we leverage the interpretability of decision trees to find the conditions in the attributes that make each data item anomalous. The path from the root to the leaf gives these conditions in the decision trees. This is expected to be useful for network administrators to identify what is causing the anomaly and allow for a fast diagnosis and solution of the corresponding network problem. Observe that this decision tree can also be used in the future to classify other anomalous data items. Suppose the network administrators have been able to identify the issues that made an experiment anomalous in the past. In that case, they can use that experience with new instances, and very possibly fix the problem quickly.

## V. Results

In this section, we report the results of our methodology, implemented by using the widely adopted scikit-learn library from Python in real data. Scikit-learn provides ML algorithms, such as decision trees and $k$-means.

### A. Data Analysis for Drive Tests

We have only chosen one value from each of categories available in the Nokia drive-test dataset. This suffices to showcase how our methodology works without losing focus in the description of experiments, and to demonstrate the properties of our proposal. The datasets used correspond to Hypertext Transfer Protocol (HTTP) file downloads, in cities, over LTE networks, and with one single operator (we anonymize the operator's name for privacy purposes). Filtering the data with these parameters produced 1494 samples. Henceforth, these are the data items (samples, experiments) to be used and analyzed in this subsection.

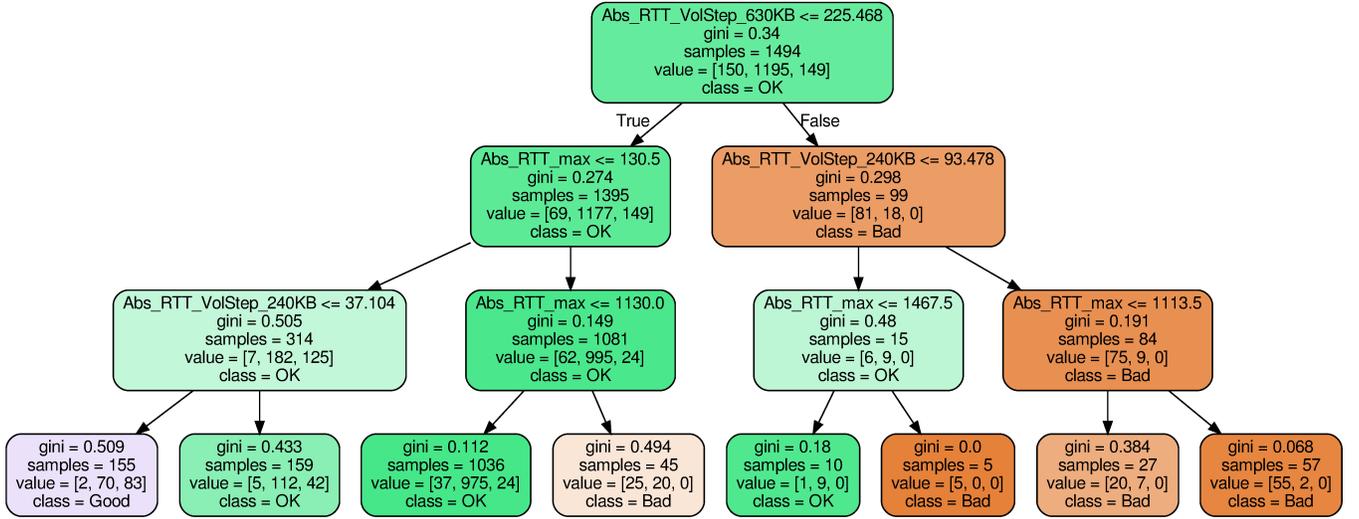

Fig. 4: Decision tree generated with supervised ML using RTT attributes as input and TDR classes inferred from percentiles (see Fig. 3). when visualizing the tree each box has the quality of the gini split, the number samples at each split, the number values for each TDR class (Bad, OK, Good), and which class was picked.

*1) TDR Characterization:* Fig. 3 depicts the distribution of the TDR values. Using these TDR values, we split the samples into three groups: those with TDR values above the 90-th percentile (Good throughput samples), below the 10-th percentile (Bad samples), and everything else (OK samples). Specifically, the value of the 10-th percentile was 7 796 kbit/s, while the value of the 90-th percentile was 68 493 kbit/s. This approach based on the use of statistical percentiles is commonly adopted by regulators and (self-) quality assessment teams for the analysis of complex systems, and by the Nokia team involved in the measurements.

*2) Detecting Anomalies:* We use the TDR split of the 1 494 samples as a target for a supervised ML decision tree with three classes: Bad, OK, and Good. Since it is well known that the TCP throughput is a function of the RTT, in the classification we use only the RTT attributes available in the dataset to feed the CART algorithm. To avoid overfitting we limited the number of internal tree levels to three. Fig. 4 illustrates the resulting decision tree with RTT attributes. For a given sample, at the root of the node, it is decided if the Abs_RTT_VolStep_630KB attribute (the RTT after downloading 630 KB) value of the sample is smaller or equal to 255.468 ms. If true, we move to the left side of the tree, and to the right side otherwise. Other information included in each tree node besides the attribute and split value is the actual value of the Gini impurity for the class split at this level of the tree, the total number of samples considered in the node (1 494 at the root) and the number of samples for each label [150, 1195, 149], that correspond to the classes [Bad, OK, Good]. The class value matches the predicted class at this level of the tree (the class that has a highest number of samples). Employing a tree traversal, we will reach a leaf node, where there are no more conditions, and the class for the sample is defined. The decision tree depicted in Fig. 4 was able to classify the TDR samples to some degree, with an accuracy score of 85% (number of correct predictions from the TDR classes computed by the tree, in this case 1 269, over the total number of samples, which is 1 494). Table II describes the results generated by the decision tree at relevant nodes.

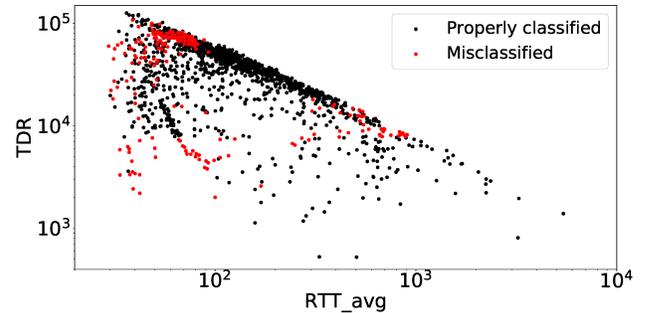

Fig. 5: Graphical plot of the data items properly classified by the decision tree of Fig. 4 in black, and the misclassified data items in red.

After using the decision tree built with the CART algorithm for the classification of all the samples in the dataset, we still have 225 samples, which represent 15% of the original dataset, that are not correctly classified (they are visually reported in Fig. 5). More in detail, Table III reports the confusion matrix, which shows how data samples labeled by means of percentile thresholds are subsequently classified by the decision tree, based on RTT attributes only. Understanding the cause of these misclassifications is the target of the methodology presented in the previous section. We therefore consider next how to cluster these anomalous samples (i.e., the data samples that are incorrectly modeled and classified with the decision tree).

*3) Clustering The Anomalies:* To cluster the misclassified samples obtained in the previous step we have chosen *k-*

TABLE III: Confusion matrix for the supervised ML classifier of Fig. 4 trained with the classes identified using Fig. 3

|  | Bad (ML) | OK (ML) | Good (ML) |
|---|---|---|---|
| **Bad (percentile)** | 105 | 43 | 2 |
| **OK (percentile)** | 29 | 1096 | 70 |
| **Good (percentile)** | 0 | 66 | 83 |

TABLE IV: radio and TCP data attributes used by the unsupervised ML algorithm ($k$-means, with $k = 2$)

| Type | Attributes | Description |
|---|---|---|
| Radio | Start.RSSI.dBm | Received signal strength indication initial value. |
| Radio | End.RSSI.dBm | Received signal strength indication (final value). |
| Radio | Start.RSRP.dBm | Reference Signals Received Power (initial value). |
| Radio | End.RSRP.dBm | Reference Signals Received Power (final value). |
| Radio | Start.SINR.dB | Signal-to-Interference-plus-Noise Ratio (initial value). |
| Radio | End.SINR.dB | Signal-to-Interference-plus-Noise Ratio (final value). |
| TCP | Abs_CWIN_avg | Average congestion window size. |
| TCP | Abs_CWIN_max | Maximum congestion Window size. |
| TCP | Abs_RWIN_avg | Average TCP receive window. |
| TCP | Abs_RWIN_max | Maximum TCP receive window. |
| TCP | Abs_PacketLost_sum | PacketLost total value. |
| TCP | Abs_IdleTime_avg | Average gap value between consecutive TCP segments. |
| TCP | triple_dupacks_b2a | Triple duplicate ack value (server to client). |

means to use as technique, due to its suitability for a medium-sized dataset and the ability to cluster data. We fit $k$-means with attributes picked from a homogeneous class of data attributes. In particular, we use either TCP-related or radio-related attributes. We try to identify if the problem belongs to TCP events (losses, duplicated ACKs, etc.,), radio quality events (e.g., changes in signal strength), or a combination of both. Table IV displays the data attributes used.

Although not presented here, we have tested various possibilities for the numbers of target clusters, and found out that the highest score of $k$-means was obtained by using only two clusters. Indeed, the use of $k$-means revealed that the data incorrectly classified by the decision tree can be further clustered into two groups according to TCP attributes. Similarly, the best choice is to use two clusters also in case of using radio attributes. Fig. 6 and Fig. 7 show the clusters obtained by using carefully chosen pairs of attributes. The first figure reports an example of TCP attributes and the second depicts an example of radio attributes. In both cases, it is clear that the clusters obtained with $k$-means separate the samples into those that have TCP (resp., radio) issues and those that do not have issues. Therefore, applying $k$-means to TCP (resp., radio) attributes allows to identify whether there exists a problem with the TCP (resp., radio) performance.

*4) Classifying Anomalies:* The last stage of our methodology involves deriving a final model, using again a decision tree, to identify the root cause of the identified anomalies. For our specific case study, we use a decision tree to classify into the cases that the causes of the anomaly are TCP events, radio conditions, or both. This decision tree has been trained with the $k$-means labels previously collected from TCP-based and radio-based clustering, and a new collection of relevant attributes (see Table IV). Fig. 8 showcases the resulting decision tree, which not only classifies TCP and radio issues, but also identifies a class of anomalies that cannot be explained by means of TCP and radio attributes (labeled as "failure to identify"). A problem is classified as "unknow" when it is not possible to distinguish between TCP or radio problem, but it is definitely one of these two. An enumeration of the rules applied by the decision tree can be found in Table V. Fig. 9 depicts the points that belong to each class in terms of TDR and RTT. Observing the final decision tree, we can distinguish if the problem is due to TCP with a probability of 0.623, or radio with probability of 0.60 (note that events are not mutually exclusive so that their probabilities do not need to sum up to one or less). The overall score of the tree is 70%, which means that 70% of the anomalies are identified, jointly with their root causes. Misclassification is mostly due to lack of data, since after three split levels, the amount of data at some leaf nodes is no longer statistically relevant. Note that the "Failure to Identify" class shown in the decision tree of Fig. 8 and in Table V is due to a lack of training data over other attributes. For example, the availability of Domain Name System (DNS) and Transport Layer Security (TLS) measurements might help to identify more types of network problems and further complement the analysis in a fully automated way.

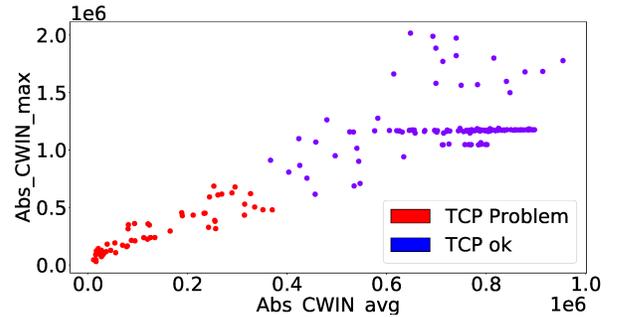

Fig. 6: The two $k$-means clusters obtained with TCP data attributes with respect to the congestion window average and maximum value attributes.

### B. MONROE Dataset

Tstat comes in as a layer to extend and validate our methodology. We choose multiple data-sets with multiple services such as Facebook and Google collected in various European countries ranging from south (Spain, Italy) to north (Sweden, Norway) using a range of mobile operators, thanks to the MONROE platform. The methodology is the same with minor tweaks; for instance, we change the TDR characterization from 90-10 to 80-20 percentiles to validate the claim that choosing other percentile thresholds does not affect the proposed methodology. Thus, everything above the 80th percentile is considered Good TDR, below the 20th Bad, and in between (20th to 80th) OK TDR. We test the methodology on three operators per country for a given service. We report the results and interesting findings per service.

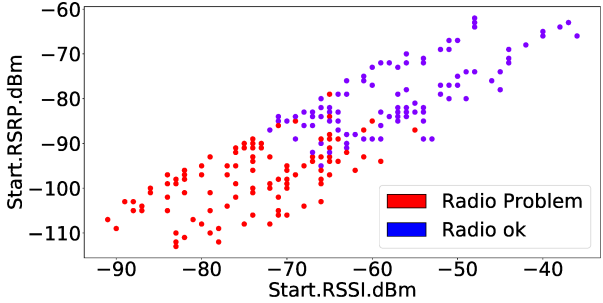

Fig. 7: The two $k$-means clusters obtained with radio attributes in Fig. 8 showcasing the dominant attributes from Table IV and classes from Fig. 8

TABLE V: Highlights of the decision tree rules for detection of anomalies in Fig. 8 showcasing the dominant attributes from Table IV and classes from Fig. 8

| Rules | Probability | Class |
|---|---|---|
| If Abs_CWIN_max <= 681468.0 | 0.54 | Radio Problem |
| If Abs_CWIN_max <= 681468.0 and if Start.RSRP.dBm <= -102.0 | x | Failure to identify (lack of data) |
| If Abs_CWIN_max <= 681468.0 and if Start.RSRP.dBm > -102.0 | 0.60 | Radio Problem |
| If Abs_CWIN_max <= 681468.0 and if Start.RSRP.dBm > -102.0 and If Abs_CWIN_avg >17639.411 | 0.66 | Radio Problem |
| If Abs_CWIN_max >681468.0 | x | Unknown problem! (investigate further) |
| If Abs_CWIN_max >681468.0 and if Start.RSRP.dBm <= -86.5 | 0.62 | TCP Problem |
| If Abs_CWIN_max >681468.0 and if Start.RSRP.dBm > -86.5 | 0.67 | Unknow problem! (investigate further) |

*1) Facebook:* The first test in question is with Facebook. We conducted multiple Facebook webpage downloads in various European countries. The average downloaded webpage size of Facebook was 72 757.3 bytes and, at each experiment run, we cleaned the cache to make sure that a new complete download occurs. In some countries, such as Sweden and Norway, we had access to mobile nodes mounted on public transport vehicles. Thus, the downloaded page statistics for those countries are a mixture of statistics for static and mobile cellular nodes, hence the slightly higher radio problems which is depicted later in this subsection. As for Italy and Spain, we only report results for static cellular nodes.

The dataset consisted of 176 018 samples. As an example, Fig. 10 exhibits the TDR values distribution for the dataset obtained using an anonymised *operator 0* in Sweden. Notice that the 20-th and 80-th percentiles are defined for each studied group of data, and if we consider the entire subset of operators and countries monitored in the study, the percentile threshold values range from 500 kbit/s up to 60 Mbit/s. Our methodological procedure uses these thresholds on the TDR split with a decision tree, while fitting only the Tstat server-side RTT attributes as the input features to classify the TDR.

The decision tree depth now changes depending on how many samples are available, and for the case of *operator 0* in Sweden, the depth was $n = 6$. As a proof that our rule on the selection of $n$ works, Fig. 12 depicts the depth versus the accuracy alongside the confidence interval applying cross-validation 30 times, again for *operator 0* in Sweden. From the figure, it is clear that with a depth of 6 we gain more accuracy while we avoid overfitting. We repeat these steps, not only for one operator in Sweden but also for all operators available in the dataset.

Table VI reports results obtained for the first decision tree, accompanied by their accuracy, precision, recall, and F1 score. To get the precision, the recall, and the F1 Score, we split the datasets into training and test subsets, with 20% testing data.

Following the methodology described in Section IV, we identify anomalous samples and use them to fit $k$-means with clusters leveraging radio and TCP attributes. For matching, we use Tstat reported attributes on radio and TCP, such as RSSI, RSRP, and RSRQ in the case of radio and Congestion and Receiver Window in TCP. Fig. 13 takes a closer look at the clusters detected by using radio attributes, where we can see that using $k$-means produces two sets of groups; TCP follows the same trend in that regard, as shown through Fig. 14. We further summarize the results with the counts of radio and TCP problem observed in the Facebook experiments with the final step of our methodology in the topmost part of Table VIII. Depending on the country and our operator, the ratio between radio and TCP problems might vary. For instance, in Sweden and Norway we can observe more radio problems than in other countries, which is due to the fact that MONROE nodes used in those countries were not only static nodes (e.g., in labs and office) but also mobile nodes (e.g., moving with buses and trains).

We can therefore deduce that our methodology can detect anomalies, and their causes, starting from the analysis of very different data sets. Indeed, our methodology easily adapts to the structure and size of the chosen dataset. We further corroborate our intuition by reporting results on different datasets and experiment, starting with the interesting finding encountered when testing the Google service, described next.

*2) Google:* A second experiment with MONROE and Tstat data leverages the Google search engine as the downloaded webpage. We observed an average downloaded size of 17 368.8 bytes, following the same procedure as the Facebook service (repeating downloads, clearing the cache, etc..). The available TDR samples are classified as shown in Fig. 11, while the first decision tree with the TDR as classes and RTT as input features yields the results reported in Table VII. The interesting part comes in treating the anomalies samples from Table VII with $k$-means. In the radio case shown in Fig. 15, $k$-means produced two sets of clusters separated, which is consistent with what previously observed in this paper with other datasets. However, a neat change comes in the TCP case where $k$-means could not identify two clusters. This is logical and somehow expected given (*i*) the small size of downloaded webpage, and (*ii*) the fact that Google services are replicated very close to users. Indeed, moving forward to our methodology's final step, we can see from Table VIII that the number of TCP problems in the Google experiment is low or non-existent; instead, we have mainly radio problems, which do not directly depend on the nature of the downloaded page and its location. This indirectly confirms that our methodology finds meaningful conclusions on the origins and causes of anomalies.

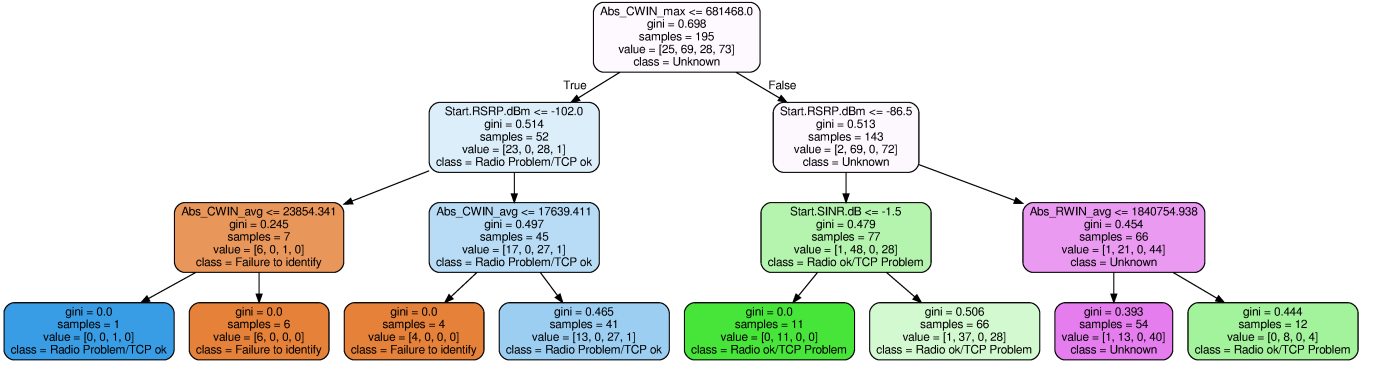

Fig. 8: Anomaly detection decision tree with TCP, radio attributes from Table IV, and *k*-means labeled clusters as classes. when visualizing the tree each box has the quality of the gini split, the number samples at each split, the number values for each class (Failure to identify, Radio Ok/TCP Problem, Radio Problem/TCP Ok, unknown), and which class was picked during each split.

TABLE VI: A summary of the initial decision tree's performance utilizing Facebook as a service in all the countries part of the MONROE project with all their operators.

| Services | Country | Operators | Accuracy (%) | Precision (%) | Recall (%) | F1 Score | # Number of samples | # Anomalies samples |
|---|---|---|---|---|---|---|---|---|
| **Facebook** | **Sweden** | *op0_sw* | 68.1 | 66.0 | 67.8 | 64.3 | 176018 | 55784 |
| | | *op1_sw* | 66.0 | 61.7 | 64.8 | 59.0 | 137777 | 46814 |
| | | *op2_sw* | 70.0 | 67.0 | 68.6 | 65.4 | 132802 | 40841 |
| | **Norway** | *op0_no* | 73.1 | 71.9 | 72.2 | 71.5 | 111276 | 29904 |
| | | *op1_no* | 74.5 | 74.4 | 74.4 | 73.4 | 132128 | 33233 |
| | | *op2_no* | 73.2 | 72.1 | 72.7 | 71.3 | 112728 | 29611 |
| | **Italy** | *op0_it* | 75.0 | 73.0 | 73.3 | 71.0 | 48265 | 12117 |
| | | *op1_it* | 75.6 | 75.4 | 75.3 | 74.7 | 73352 | 17869 |
| | | *op2_it* | 70.6 | 67.3 | 67.6 | 66.4 | 36223 | 10622 |
| | **Spain** | *op0_es* | 74.7 | 73.6 | 73.4 | 71.3 | 27289 | 7025 |
| | | *op1_es* | 76.4 | 75.5 | 76.2 | 75.2 | 75903 | 17878 |
| | | *op2_es* | 72.5 | 70.9 | 70.7 | 67.8 | 23449 | 6442 |

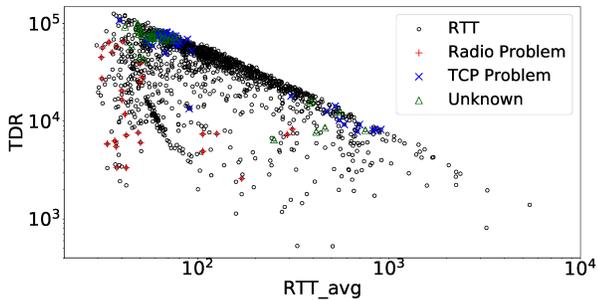

Fig. 9: The outcome of the misclassified points using a combination of unsupervised and supervised ML in Fig. 8 and properly classified points with the supervised ML classifier in Fig. 4.

*3) Other services:* We have used our methodology to analyze more datasets collected with MONROE and Tstat, e.g., by repeatedly accessing the webpages of Youtube (80 kbytes on average) and Twitter (39.1 kbytes on average). Youtube showed the same behavior as Google, with almost no TCP problem, while the analysis with Twitter's experiments was similar to the one with Facebook. Table VIII reports the final results obtained in the four cases by applying our methodology. These results were expected because YouTube and Google are typically co-deployed, while Facebook and Twitter have similar webpage structures. They further confirm that our methodology yields consistent results across different experimental conditions.

Regardless of the service, country, or operator, with this extended version, we have shown that our automated anomaly detection system works and is functional to improve the networking service quality. Indeed, the system can self-identify network performance issues. So it can be used to alert the right support personnel, either TCP or radio experts in the presented example, which will receive as input the rules from, e.g., Table V that raised the alarm. Our methodology is easy to implement, and it can be deployed in production environments.

## VI. CONCLUSIONS

In this paper, we have presented a methodology that allows to fully automate the process of identifying anomalies in the behavior of a network. The main advantages of the proposed approach are that it can be implemented in a fully automated way and that its results are interpretable by humans. Specifically, our methodology is based on the combined application of well-established supervised and unsupervised ML tools.

We have also provided a few application examples based on real data from operational cellular networks. Notwithstanding

TABLE VII: A summary of the initial decision tree's performance utilizing Google as a service following up the same methodology steps from section V-B1

| Services | Country | Operators | Accuracy (%) | Precision (%) | Recall (%) | F1 Score (%) | # Number of samples | # Anomalies samples |
|---|---|---|---|---|---|---|---|---|
| Google | Sweden | $op0\_sw$ | 64.8 | 66.0 | 67.7 | 65.4 | 100876 | 35475 |
| | | $op1\_sw$ | 66.0 | 66.0 | 67.7 | 65.5 | 97945 | 34760 |
| | | $op2\_sw$ | 70.0 | 66.8 | 68.6 | 64.6 | 90198 | 31550 |
| | Norway | $op0\_no$ | 66.9 | 70.2 | 69.9 | 66.3 | 52158 | 17228 |
| | | $op1\_no$ | 67.6 | 71.4 | 70.3 | 67.8 | 42028 | 13586 |
| | | $op2\_no$ | 66.8 | 69.7 | 69.9 | 67.6 | 54996 | 18261 |
| | Italy | $op0\_it$ | 67.7 | 72.1 | 71.2 | 68.9 | 74596 | 24091 |
| | | $op1\_it$ | 66.5 | 69.0 | 69.0 | 66.2 | 46978 | 15726 |
| | | $op2\_it$ | 66.0 | 69.3 | 67.6 | 66.4 | 32909 | 11164 |
| | Spain | $op0\_es$ | 69 | 71.9 | 71.7 | 70.5 | 75279 | 23334 |
| | | $op1\_es$ | 66.0 | 70.5 | 69.7 | 66.6 | 31863 | 10821 |
| | | $op2\_es$ | 64.8 | 68.4 | 68.3 | 67.8 | 27058 | 8315 |

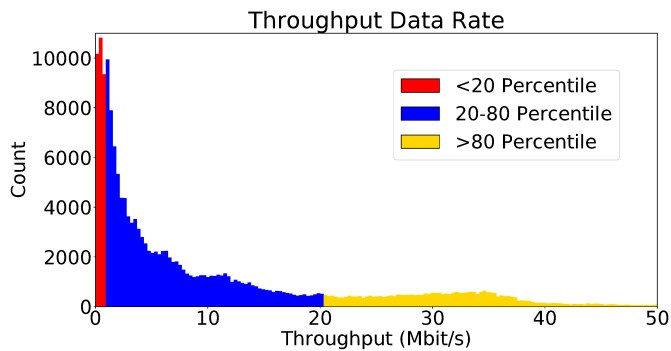

Fig. 10: A snapshot of the Distribution of Throughput Data Rate in Sweden using Operator 0 and Facebook as a service.

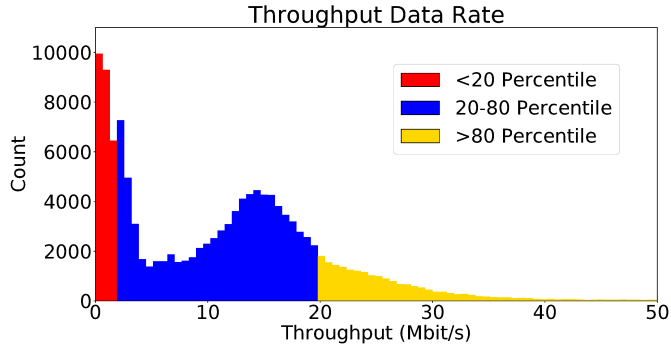

Fig. 11: A snapshot of the Distribution of Throughput Data Rate in Sweden using Operator 0 and Google as a service.

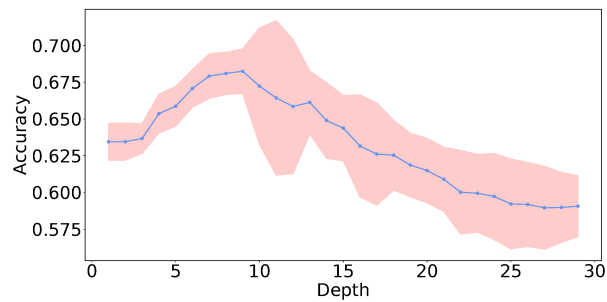

Fig. 12: The depth versus accuracy for a decision tree with the confidence interval using Facebook as service in Sweden with Operator 0.

### ACKNOWLEDGMENTS

This research was supported by the Region of Madrid through the TAPIR-CM project (S2018/TCS-4496). The work of V. Mancuso was supported by the Ramon y Cajal grant RYC-2014-16285 from the Spanish Ministry of Economy and Competitiveness.

the datasets were obtained under very heterogeneous conditions and from very different networks, we have shown that we can easily identify behavioral anomalies (specifically, we have dealt with the case of TCP throughput, although the methodology is not tied to TCP) and that we are also able to further investigate and identify the their root causes. This feature is key to promptly activate precise and effective troubleshooting actions.

Furthermore, our methodology is generic and can be used to automatically detect several types of networking problems, and examine several classes of potential anomaly causes, without losing the interpretation of the results. As such, it can be stationed by current cellular networks and used for future cellular monitoring infrastructures.

TABLE VIII: The final decision tree performance across countries in the MONROE project alongside their operators using the samples trained by $k$-means from Fig. 13 and Fig. 14 for radio, and TCP respectively.

| Services | Country | Operators | Accuracy (%) | Precision (%) | Recall (%) | F1 Score | Radio Problem | TCP Problem | Prob in both |
|---|---|---|---|---|---|---|---|---|---|
| Facebook | Sweden | op0_sw | 99.1 | 98.0 | 96.0 | 98 | 13082 | 5153 | 2963 |
| | | op1_sw | 98.7 | 97.2 | 95.2 | 96.1 | 5914 | 8065 | 5311 |
| | | op2_sw | 99.0 | 98.2 | 97.4 | 98.3 | 19416 | 2487 | 4416 |
| | Norway | op0_no | 99.3 | 99.2 | 98.7 | 98.9 | 10747 | 3188 | 2990 |
| | | op1_no | 99.3 | 99.2 | 98.3 | 98.0 | 12136 | 2384 | 2877 |
| | | op2_no | 99.0 | 98.0 | 97.0 | 97.0 | 3304 | 11924 | 9797 |
| | Italy | op0_it | 99.4 | 98.4 | 97.2 | 98.0 | 1642 | 6750 | 4425 |
| | | op1_it | 98.7 | 92.6 | 94.4 | 92.5 | 801 | 2880 | 5194 |
| | | op2_it | 99.5 | 99.0 | 99.0 | 99.0 | 372 | 5357 | 1884 |
| | Spain | op0_es | 98.3 | 98.5 | 97.0 | 96.0 | 2631 | 5039 | 9008 |
| | | op1_es | 99.3 | 98.2 | 98.1 | 98.0 | 656 | 3544 | 2314 |
| | | op2_es | 99.0 | 98.3 | 98.5 | 97.7 | 1046 | 1057 | 4493 |
| Google | Sweden | op0_sw | 99.8 | 99.9 | 99.9 | 99.9 | 11120 | 1 | 122 |
| | | op1_sw | 99.7 | 99.7 | 99.7 | 99.7 | 22799 | 2 | 203 |
| | | op2_sw | 99.4 | 99.6 | 99.6 | 99.4 | 9653 | 3 | 106 |
| | Norway | op0_no | 99.3 | 99.5 | 99.5 | 99.5 | 4852 | 681 | 637 |
| | | op1_no | 99.4 | 99.7 | 99.7 | 99.7 | 7839 | 8 | 5 |
| | | op2_no | 98.5 | 98.7 | 98.7 | 98.7 | 10591 | 333 | 422 |
| | Italy | op0_it | 99.3 | 99.5 | 99.5 | 99.5 | 12121 | 69 | 23 |
| | | op1_it | 99.7 | 99.7 | 99.7 | 99.7 | 7141 | 118 | 71 |
| | | op2_it | 98.4 | 99.2 | 99.2 | 99.2 | 4591 | 476 | 372 |
| | Spain | op0_es | 98.7 | 98.5 | 98.9 | 98.9 | 14058 | 30 | 716 |
| | | op1_es | 98.7 | 98.9 | 99.0 | 99.0 | 4227 | 23 | - |
| | | op2_es | 99.0 | 99.7 | 99.7 | 99.7 | 1586 | 8 | 2 |
| Youtube | Sweden | op0_sw | 98.8 | 98.9 | 98.9 | 98.9 | 121200 | 2340 | 122 |
| | | op1_sw | 98.8 | 98.9 | 98.9 | 98.9 | 92473 | 897 | 449 |
| | | op2_sw | 99.1 | 99.2 | 99.2 | 99.2 | 80054 | 4280 | 1203 |
| | Norway | op0_no | 99.6 | 99.6 | 99.6 | 99.6 | 171094 | 534 | 1814 |
| | | op1_no | 99.3 | 99.3 | 99.3 | 99.3 | 163712 | 234 | 7567 |
| | | op2_no | 99.5 | 99.5 | 99.5 | 99.5 | 61605 | 1960 | 26 |
| | Italy | op0_it | 99.3 | 99.3 | 99.3 | 99.3 | 47544 | 28 | 268 |
| | | op1_it | 99.5 | 99.5 | 99.5 | 99.5 | 22503 | 3071 | 1071 |
| | | op2_it | 97.9 | 98.0 | 98.0 | 98.0 | 18266 | 2155 | 804 |
| | Spain | op0_es | 98.7 | 98.5 | 98.9 | 98.9 | 80248 | 2 | 20 |
| | | op1_es | 98.7 | 98.9 | 99.0 | 99.0 | 28229 | 1884 | 2951 |
| | | op2_es | 99.0 | 99.7 | 99.7 | 99.7 | 22430 | 840 | 1020 |
| Twitter | Sweden | op0_sw | 98.8 | 98.9 | 98.9 | 98.9 | 23591 | 15165 | 6565 |
| | | op1_sw | 98.8 | 98.9 | 98.9 | 98.9 | 24907 | 7442 | 2653 |
| | | op2_sw | 99.1 | 99.2 | 99.2 | 99.2 | 37615 | 11108 | 18926 |
| | Norway | op0_no | 98.8 | 98.7 | 98.7 | 98.7 | 66112 | 4306 | 6803 |
| | | op1_no | 98.3 | 98.3 | 98.3 | 98.3 | 34017 | 14908 | 19618 |
| | | op2_no | 98.7 | 98.7 | 98.7 | 98.7 | 17406 | 23543 | 28127 |
| | Italy | op0_it | 99.3 | 99.5 | 99.5 | 99.5 | 27306 | 3051 | 2894 |
| | | op1_it | 99.7 | 99.7 | 99.7 | 99.7 | 14679 | 7584 | 2674 |
| | | op2_it | 98.4 | 99.2 | 99.2 | 99.2 | 17378 | 14 | 6 |
| | Spain | op0_es | 96.7 | 96.7 | 96.6 | 96.6 | 19041 | 8280 | 7896 |
| | | op1_es | 98.9 | 98.8 | 98.8 | 98.8 | 5403 | 1364 | 987 |
| | | op2_es | 99.0 | 98.8 | 99.8 | 99.8 | 8459 | 2572 | 4288 |

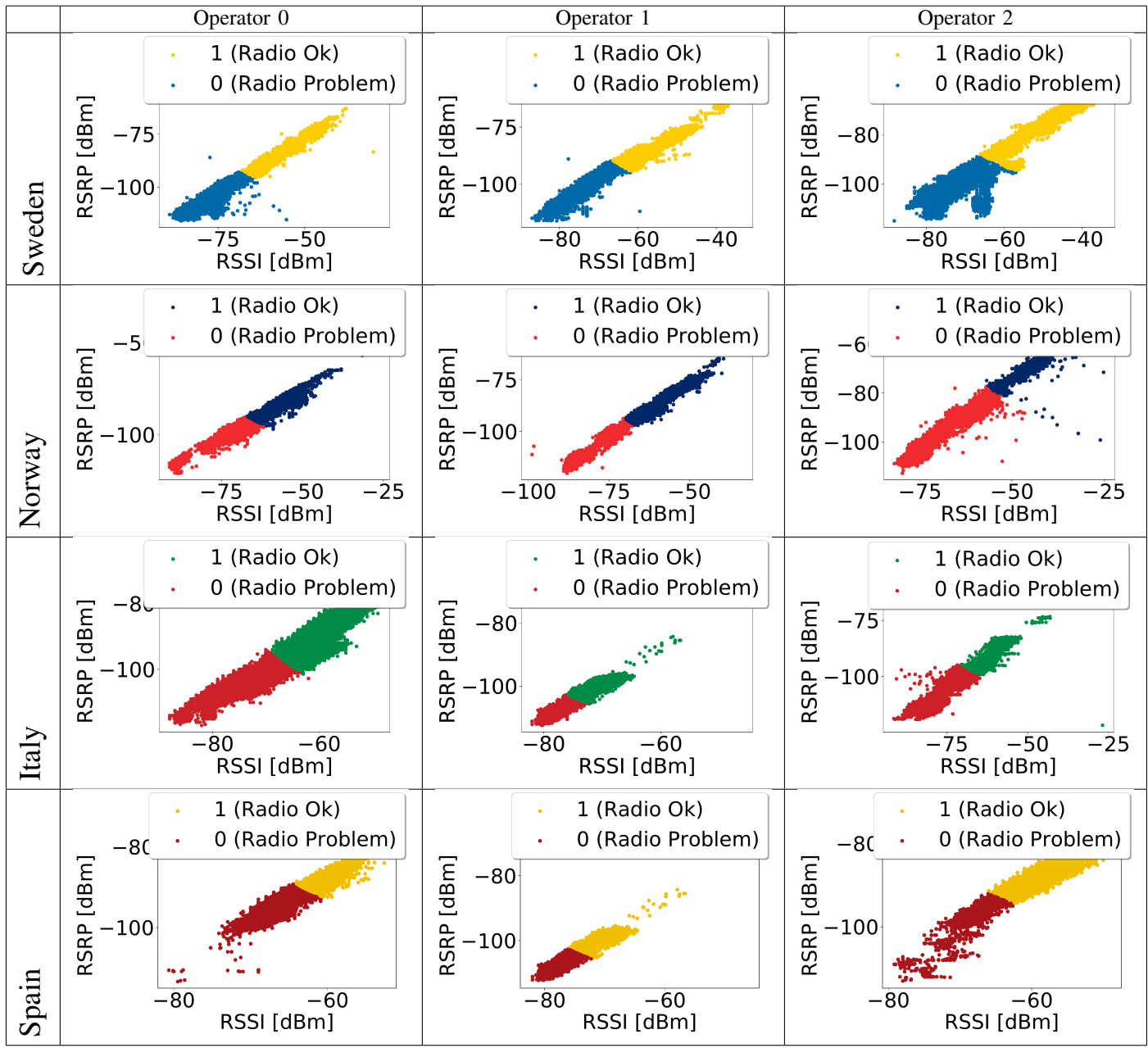

Fig. 13: The two *k*-means clusters per countries and operators obtained with radio attributes concerning the RSSI and RSRP attributes using anomalies sample from table VI leveraging Facebook as a service.

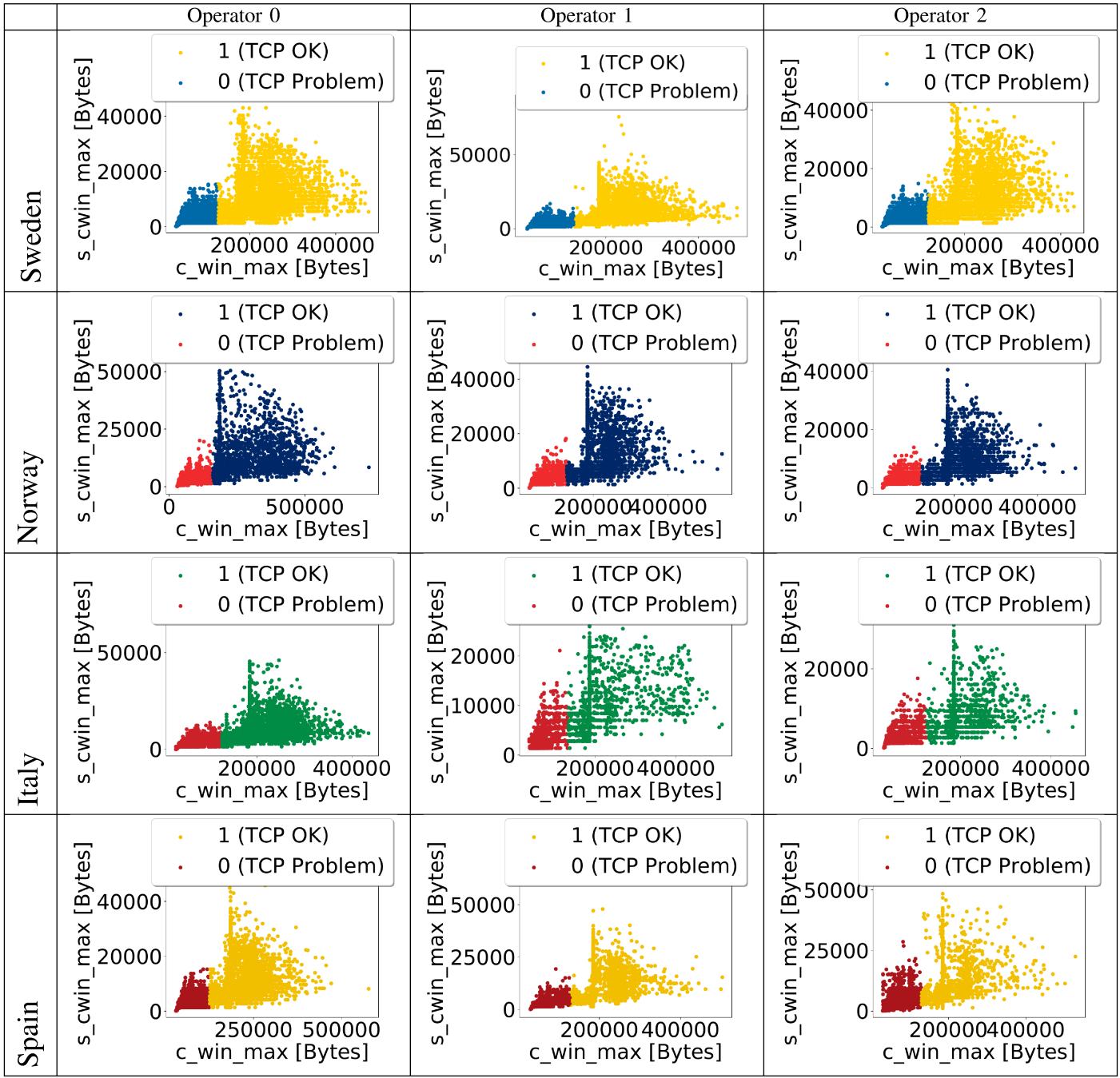

Fig. 14: The two *k*-means clusters per countries and operators obtained with TCP attributes using anomalies sample from table VI leveraging Facebook as a service

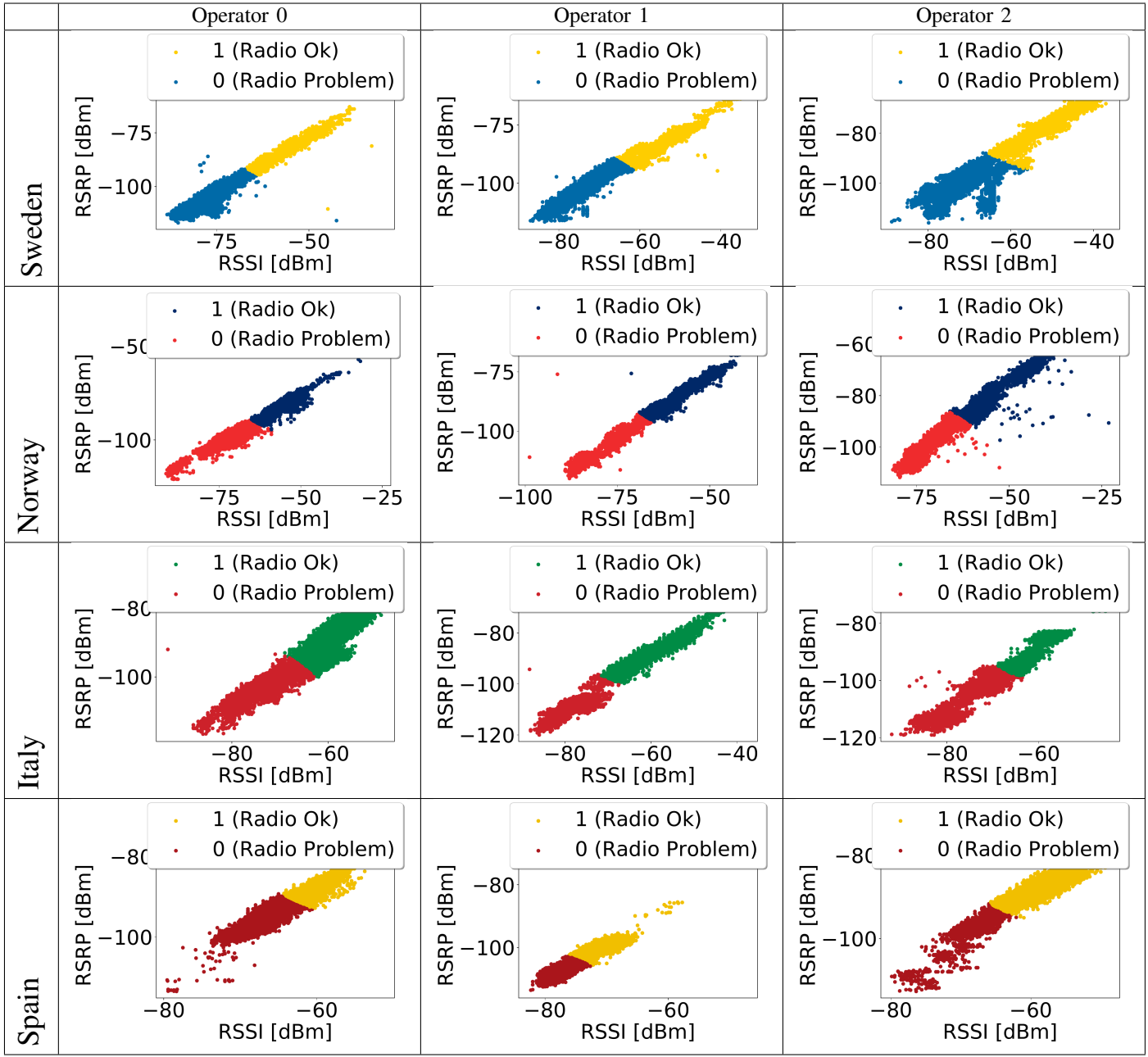

Fig. 15: The two *k*-means clusters per countries and operators obtained with radio attributes concerning the RSSI and RSRP attributes using anomalies sample from table VII leveraging google as a service.

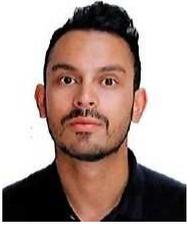

**Mohamed Moulay** is a PhD student at the Universidad Carlos III de Madrid, Spain. He received his B.Sc degree in electronics and communication engineering from the Applied Science University in Amman, Jordan in 2016 and his M.Sc. degree in multimedia and communications in 2017 from the Universidad Carlos III de Madrid, Spain. Previously, he was with IMDEA Networks as a Research Engineer. His current research interest focuses on detecting and resolving networking problems using Machine Learning algorithms.

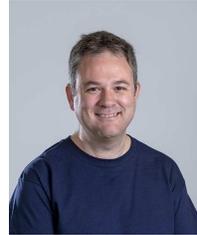

**Rafael Garcia Leiva** has a Bachelors degree in Computer Science and a Master degree in Computational Sciences. He worked for four years at University of Córdoba as a scientific programmer. He worked for three years at CERN-UAM as research engineer in high performance computing. He worked for three years as R&D Director at Andago Ingeniería in the area of open source software. In 2008 he funded Entropy Computational Services where he worked for five years in cloud computing and quantitative trading. In 2014 he joined to the Madrid Institute for Advanced Studies as research engineer in artificial intelligence.

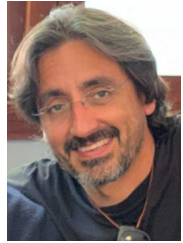

**Pablo J. Rojo Maroni** Distinguished Member of the Technical Staff at Nokia where he has been working 20 years as a E2E Network Optimization Consultant, Data Scientist, IP Solution Architect and System Architect in a wide variety of Telecommunication Mobile Network projects around the world. Currently, he is a Phd student at the Universidad Carlos III de Madrid, Spain focused in the research of new massive data processing tools, techniques and algorithms for mobile networks applications performance analysis and optimization.

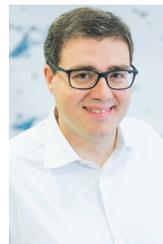

**Vincenzo Mancuso** is Research Associate Professor at IMDEA Networks, Madrid, Spain, and recipient of a Ramon y Cajal research grant of the Spanish Ministry of Science and Innovation. Previously, he was with INRIA (France), Rice University (USA) and University of Palermo (Italy), from where he obtained his Ph.D. in 2005. His research focus is on analysis, design, and experimental evaluation of opportunistic wireless architectures and mobile broadband services.

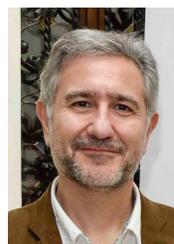

**Antonio Fernández Anta** is Research Professor at IMDEA Networks. Previously he was a on the Faculty of the Universidad Rey Juan Carlos (URJC), and the Universidad Politécnica de Madrid (UPM), where he received a research performance award. He was a postdoc at MIT (1995-1997), and spent sabbatical years at Bell Labs and MIT Media Lab. He has been awarded the Premio Nacional de Informática "Aritmel" in 2019 and is Mercator Fellow of the SFB MAKI in Germany since 2018. He received his M.Sc. and Ph.D. from the University of Louisiana. He is a Senior Member of ACM and IEEE.

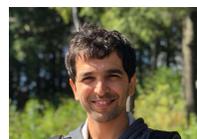

**Ali Safari Khatouni** received his B.S. degree in Software Engineering from Urmia University, Iran, and M.S. and PhD degrees from the Department of Electrical and Computer Engineering at Politecnico di Torino, Italy. Currently, he is a data scientist at Shopify. He was a postdoctoral fellow at the Faculty of Computer Science at Dalhousie University and Western University. His research interests lie in the areas of data science, network traffic analysis, machine learning, and mobile broadband networks. Moreover, He has obtained valuable experience in several European research projects (Mplane, MONROE).